\documentclass[aps,prb,twocolumn,showpacs]{revtex4}

\usepackage{graphicx}
\def  \bsig    {\mbox{\boldmath$\sigma$}}

\begin{document}

\title{Anomalous Hall effect in a two-dimensional electron gas
with spin-orbit interaction}
\author{V.~K.~Dugaev$^{1,2,\dag }$, P. Bruno$^1$, M. Taillefumier$^{1,3}$,
B. Canals$^3$, and C. Lacroix$^3$}
\affiliation{$^1$Max-Planck-Institut f\"ur Mikrostrukturphysik,
Weinberg 2, 06120 Halle, Germany\\
$^2$Department of Physics and CFIF, Instituto Superior T\'ecnico,
Av. Rovisco Pais, 1049-001 Lisbon, Portugal\\
$^3$Laboratoire Louis N\'eel, CNRS, Boite Postale 166, 38042
Grenoble Cedex 09, France}
\date{\today }

\begin{abstract}
We discuss the mechanism of anomalous Hall effect related to the
contribution of electron states below the Fermi surface (induced
by the Berry phase in momentum space). Our main calculations are made
within a model of two-dimensional electron gas with spin-orbit
interaction of the Rashba type, taking into account the
scattering from impurities. We demonstrate that such an
"intrinsic" mechanism can dominate but there is a competition with
the impurity-scattering mechanism, related to the contribution of
states in the vicinity of Fermi surface. We also show that the
contribution to the Hall conductivity from electron states close
to the Fermi surface has the intrinsic properties as well.
\vskip0.5cm \noindent
\end{abstract}
\pacs{73.20.Fz; 72.15.Rn; 72.10.Fk}
\maketitle

\section{Introduction}

The measurement of anomalous Hall effect\cite{hurd} (AHE) in
ferromagnetic metals and semiconductors is a powerful tool to
characterize the magnetization of thin films and mesoscopic
objects. That is why it attracted recently a lot of attention in
the spintronics community.\cite{ohno98} On the other hand, the
theoretical aspects and possible physical mechanisms of AHE are
still not quite
well understood, and the great efforts are directed now toward the
physical explanation of AHE in magnetic semiconductors, pyroclore
crystals, and magnetic compounds with inhomogeneous magnetization.
The state of art in the theory of AHE has been recently reviewed
by Sinova {\it et al}.\cite{sinova04}

Relatively well studied interpretations of AHE include the
skew scattering \cite{karplus54,smit58} and side-jump
\cite{berger70} mechanisms, which have been recently revised and
physically clarified by Cr\'epieux {\it et
al}.\cite{crepieux01,dugaev01} These mechanisms are related to the
scattering from impurities, with the spin-orbit (SO) interaction
playing the leading role, so that the spin-flip interaction is
included into the amplitude of scattering from impurities.

It was shown recently that the nonzero spin chirality in some
inhomogeneous ferromagnets also leads to the nonvanishing AHE.
\cite{matl98,ye99,chun00,taguchi01,tatara02} Such a {\it chirality}
mechanism of AHE is related to the Berry phase \cite{berry84}
acquired by the wave function of electrons moving adiabatically in
the inhomogeneous magnetization field. In principle, the SO
interaction is not a necessary element of this mechanism.
However, in earlier presented theoretical models, the
nonvanishing average AHE was obtained only in the presence of SO
interaction. Otherwise, the spatially averaged off-diagonal
conductivity was found to be zero.
A model of magnetic nanostructure without SO interaction, leading
to the Berry phase induced AHE, \cite{bruno04} has been called the {\it
topological} mechanism of AHE.

Now the main interest is directed toward another {\it
"intrinsic"} mechanism of AHE recently proposed by MacDonald {\it et al.}
\cite{jungwirth02,yao04} and Nagaosa {\it et al.} \cite{m_onoda02}
This mechanism uses the SO interaction included into the
Hamiltonian of electrons in the crystal lattice.
The key element of this theory is the Berry phase
in the {\it momentum} space, so that a nontrivial topology of the
electron energy bands is necessary for the intrinsic mechanism. A
similar idea of the calculations of Hall conductivity has been
used long ago in the context of quantum Hall effect.
\cite{thouless82,volovik88}

The Berry-phase-induced intrinsic mechanism of AHE does not take
into account scattering from impurities. It was suggested before
that the impurities can be totally neglected in case of intrinsic
SO interaction because the effect of impurities should be
necessarily extrinsic. Correspondingly, the calculations of AHE
have been performed in a zero density-of-impurity limit, $N_i\to
0$,\cite{jungwirth02,m_onoda02} which was taken before the static
limit of $\omega \to 0$ in the Kubo formula for conductivity
$\sigma _{ij}(\omega )$.

In this paper we present the result of calculations taking into
account the effect of impurity scattering. We demonstrate that
there exists another intrinsic contribution to AHE related to the
electron states near the Fermi surface. This contribution is not
geometric, and we show that it can be of the same order of
magnitude as the Berry-phase-induced one, which has been
calculated before. To calculate properly the non-geometric
contribution, we should take into account scattering from
impurities leading to the finite relaxation time of electron
states at the Fermi surface. Accordingly, we take the limit of
$\omega \to 0$ {\it  before} $N_i\to 0$, which ensures the
transition between disordered and clean to be continuous. At the
same time, the non-geometric contribution is intrinsic, and it
does not vanish if we finally take the limit of $N_i\to 0$.

In our calculations, we use a model of two-dimensional electron
gas (2DEG) with SO interaction in the form of Rashba
term.\cite{rashba} We show that, in accordance with the well-known
result of St$\check{\rm r}$eda,
\cite{streda82}$^,$\cite{crepieux01} there are two terms in the
off-diagonal conductivity, one of which, $\sigma _{xy}^I\, $, is
due to the electron states near the Fermi energy (it corresponds
to the non-geometric contribution), and the other one, $\sigma
_{xy}^{II}\, $, is related to the contribution of all occupied
electron states below the Fermi energy (identified as the
Berry-phase-induced intrinsic mechanism, which has been calculated
before).

In agreement with other works on different models of the electron
energy spectrum, our calculation confirm that for the 2DEG with
Rashba SO interaction, there is a non-vanishing contribution
$\sigma ^{II}_{xy}$ from the states below the Fermi surface. This
contribution is absent in the case of the side-jump and skew
scattering of electrons within the model of a parabolic energy
band. However, we emphasize that the contribution to $\sigma
_{xy}$ from the vicinity of the Fermi surface ($\sigma ^I_{xy}$)
can not be totally neglected.

Recently, the role of impurity scattering in AHE has been analyzed
within the double-exchange model with non-coplanar spin configuration.
\cite{onoda04} It was shown that two different mechanisms can contribute
to the AHE, one of which is related to the spin texture in the real
space and the other to a momentum-space skyrmion-density at the
Fermi level. In the latter case, the impurity scattering was relevant.
This conclusion is in agreement with our results on a different model.

\section{Model and the Kubo formula}

We consider a model of 2DEG in the $x-y$ plane under a homogeneous
magnetization field ${\bf M}_0$ directed along the axis $z$, with
the SO interaction described by the Rashba term.\cite{rashba} The
Hamiltonian of this model reads
\begin{equation}
\label{1}
H_0=\varepsilon _{\bf k}
+\alpha \left( \sigma _x\, k_y-\sigma _y\, k_x\right)
-M\sigma _z\; ,
\end{equation}
where $\varepsilon _{\bf k}=k^2/2m$, $\alpha $ is the coupling
constant of SO interaction, $M=g\mu _BM_0$, $M_0$ is the magnitude
of magnetization, and we take units with $\hbar =1$. Due to the SO
interaction, the components of velocity operators, $v_i =\partial
H/\partial k_i$, are matrices in the spin space
\begin{equation}
\label{2}
v_x=\frac{k_x}{m}-\alpha \, \sigma _y,\hskip0.5cm
v_y=\frac{k_y}{m}+\alpha \, \sigma _x .
\end{equation}

We also include into consideration the scattering from impurities,
described by a disorder potential $V({\bf r})$. We assume that the
disorder potential is short-range and weak, and we will treat it
in the Born approximation of the impurity scattering. Thus, the
Hamiltonian of our model is $H=H_0+V({\bf r})$, and any physical
variables including the Hall conductivity should be calculated with a
corresponding averaging over the random potential $V({\bf r})$.

It should be noted that such a model does not take into account
the SO interaction in the impurity potential $V({\bf r})$. It is
known that the SO interaction in the impurity scattering produces
the AHE {\it via} well-studied side-jump and skew scattering
mechanisms (see Ref.~[\onlinecite{crepieux01}] and references
therein). Here we concentrate on other possible mechanisms induced
by the SO interaction in the potential of the crystal lattice.

The conductivity tensor can be calculated using the general
Kubo formalism
\begin{equation}
\label{3}
\sigma _{ij}(\omega )
=\frac{e^2}{\omega }\, {\rm Tr}
\int \frac{d\varepsilon }{2\pi }\;
\left< \hat{v}_i\, \hat{G}(\varepsilon +\omega )\,
\hat{v}_j\, \hat{G}(\varepsilon )\right> ,
\end{equation}
where $\hat{G}(\varepsilon )=\left( \varepsilon -H\right) ^{-1}$
is the operator Green function of electrons described by the
Hamiltonian $H$, which includes the disorder, $\hat{v}_i$ is the
corresponding velocity operator, the trace goes over any
eigenstates in the space of operator $H$ (e.g., eigenstates of the
Hamiltonian $H$ itself), and $\left< ... \right> $ means the
disorder average.

It was first shown by St$\check{\rm r}$eda\cite{streda82} and
later in context of the AHE by Cr\'epieux {\it et
al.}\cite{crepieux01} that a general formula for the static
conductivity $\sigma _{ij}(\omega =0)$ can be presented in a form of
two different terms, $\sigma _{xy}=\sigma _{xy}^I+\sigma
_{xy}^{II}\, $, one of which, $\sigma _{xy}^I\, $, stems from the
contribution of electrons at the Fermi surface
\begin{eqnarray}
\label{4}
\sigma ^I_{ij}=\frac{e^2}2\, {\rm Tr}
\int \frac{d\varepsilon }{2\pi }
\left( -\frac{\partial f(\varepsilon )}{\partial \varepsilon }\right)
\left< \hat{v}_i
\left[ \hat{G}^R(\varepsilon )-\hat{G}^A(\varepsilon )\right]
\right. \nonumber \\ \left. \times \,
\hat{v}_j\, \hat{G}^A(\varepsilon )
-\hat{v}_i\, \hat{G}^R(\varepsilon )\, \hat{v}_j
\left[ \hat{G}^R(\varepsilon )-\hat{G}^A(\varepsilon )\right]
\right> ,
\end{eqnarray}
and the second one, $\sigma _{xy}^{II}\, $, formally contains the
contribution of all filled states below the Fermi energy
\begin{eqnarray}
\label{5}
\sigma ^{II}_{ij}=\frac{e^2}2\, {\rm Tr}
\int \frac{d\varepsilon }{2\pi }\; f(\varepsilon )
\left<
\hat{v}_i\;
\frac{\partial \hat{G}^A(\varepsilon )}{\partial \varepsilon }\;
\hat{v}_j\, \hat{G}^A(\varepsilon )
\right. \nonumber \\ \left.
-\, \hat{v}_i\,
\hat{G}^A(\varepsilon )\, \hat{v}_j\;
\frac{\partial \hat{G}^A(\varepsilon )}{\partial \varepsilon }\;
+\hat{v}_i\; \hat{G}^R(\varepsilon )\, \hat{v}_j\;
\frac{\partial \hat{G}^R(\varepsilon )}{\partial \varepsilon }\;
\right. \nonumber \\ \left.
-\, \hat{v}_i\,
\frac{\partial \hat{G}^R(\varepsilon )}{\partial \varepsilon }\;
\hat{v}_j\, \hat{G}^R(\varepsilon )
\right> .
\end{eqnarray}

In this paper, for the model of Eq.~(1) with the impurity disorder, using a
different way of calculation, we come to a similar result in
accordance with the St$\check{\rm r}$eda formula. We calculate the
off-diagonal conductivity $\sigma _{xy}$ in the loop approximation,
changing the disorder average in Eq.~(3) as $\left< v_x\, G\,
v_y\, G\right> \to \mathcal{T}_x\left< G\right> v_y \left< G\right> $,
where $\mathcal{T}_x$ is the renormalized velocity related to the vertex
corrections.\cite{agd}
In this approximation we do not take into account the localization
corrections,\cite{lee85} which are vanishingly small
in the limit of small concentration of impurities. The
vertex corrections are known to vanish in the case of short-range
impurity potential and simple parabolic energy spectrum.
As we show in Sec.~3, in the case of a complex band structure,
the vertex corrections do not vanish even for the short-range impurity
potential.

To simplify the notations, in the following we omit the angular brackets
for $\left< G\right> $, and we denote the disorder-averaged Green function
as $G_{\bf k}(\varepsilon )$. Here we use the momentum representation for
$G_{\bf k}(\varepsilon )$ since the disorder-averaged Green function is
diagonal in this representation.

Hence, we will calculate the off-diagonal conductivity from
\begin{equation}
\label{6}
\sigma _{xy}(\omega )
=\frac{e^2}{\omega }\, {\rm Tr}\int \frac{d\varepsilon }{2\pi }\,
\frac{d^2{\bf k}}{(2\pi )^2}\;
\mathcal{T}_x(\varepsilon ,\omega)\, G_{\bf k}(\varepsilon +\omega )\,
v_y\, G_{\bf k}(\varepsilon ) ,
\end{equation}
where the trace runs over the spin states.

In the following calculation of the static off-diagonal conductivity
$\sigma _{xy}(\omega =0)$, we will be interested in the case when the
density of impurities $N_i$ is small but finite. It also includes
the "clean limit" of $N_i\to 0$, which can be physically realized
in samples with vanishingly small concentration of impurities and defects.
Correspondingly, when we calculate the static conductivity
tensor of a clean sample using Eq.~(6), we take the limit of $\omega \to 0$
before the limit of $N_i\to 0$.

It should be noted that in all previous considerations
of the intrinsic mechanism of AHE, the impurities were totally
neglected under an assumption that the effect of impurities is always
"extrinsic". Here we emphasize that this was essentially wrong.
Performing the calculations, which include the scattering from
impurities, here we find an additional contribution, which is also
intrinsic, i.e., it does not vanish in the limit of $N_i\to 0$.
This result is quite similar to the known side-jump mechanism of AHE.
\cite{crepieux01} As we will see, the additional contribution to AHE
comes from the states at the Fermi surface, for which the finite
relaxation time due to impurity scattering is important.
If we totally neglect the scattering from impurities, we miss this term.
The real structures always have some impurities or defects. Therefore,
we can justify taking the limit of $\omega \to 0$ before $N_i\to 0$
as a physical realization of the clean system as a system with
very small but still nonzero density of impurities.

In the absence of impurity scattering, the electron Green function
can be found using Hamiltonian (1)
\begin{eqnarray}
\label{7}
G_{\bf k}^0(\varepsilon )
=\frac{\varepsilon -\varepsilon _{\bf k}+\mu
+\alpha \left( k_y\sigma _x-k_x\sigma _y\right) -M\sigma _z}
{\varepsilon -E_{{\bf k},+}+\mu
+i\delta \,{\rm sign}\, \varepsilon }
\nonumber \\
\times \,
\frac1{\varepsilon -E_{{\bf k},-}+\mu
+i\delta \,{\rm sign}\, \varepsilon }\; ,
\end{eqnarray}
where $\mu $ is the chemical potential and the electron energy
spectrum $E_{\bf k}$ consists of two branches, which we label as
"+" and "--" corresponding mostly to the spin up and down electrons,
respectively (even though they contain an admixture of the
opposite spin due to the SO interaction),
\begin{equation}
\label{8}
E_{{\bf k},\pm }=\varepsilon _{\bf k}\mp \lambda (k),
\end{equation}
and we denote $\lambda (k)=\left( M^2+\alpha ^2 k^2\right) ^{1/2}$.

We consider a general case when the chemical potential $\mu $ can be situated
in both spin up and down subbands, corresponding to $\mu >M$, as shown
schematically in Fig.~1.
When only the spin up subband is filled with electrons, $-M<\mu <M$, only
the contribution of the filled subband, $E_{{\bf k},+}$, should be hold.

\begin{figure}
\includegraphics[scale=0.4]{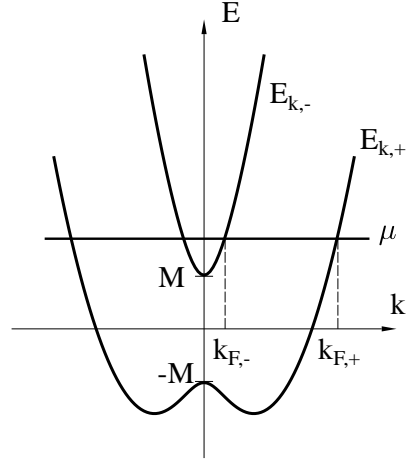}
\vspace*{-0.7cm}
\caption{Energy spectrum of electrons in a two-dimensional ferromagnet
with Rashba SO interaction (schematically).}
\end{figure}

Due to the scattering from impurities and defects, the Green function (7)
is modified. It is important to include the effect of scattering
for the correct evaluation of the contribution to the off-diagonal
conductivity from the Fermi surface.
For simplicity, we consider the model of disorder created by weak short-range
scatterers, which can be treated in the Born approximation. The corresponding
self energy of electrons is calculated as
\begin{equation}
\label{9}
\Sigma _i\, (\varepsilon )=N_i\, V_0^2
\int \frac{d^2{\bf k}}{(2\pi )^2}\; G_{\bf k}^0(\varepsilon ),
\end{equation}
where $V_0$ is the Fourier transform of the impurity potential,
and $N_i$ is the impurity density. The result of calculation
using Eqs.~(7) and (9) can be presented as
\begin{equation}
\label{10}
\Sigma _i\, (\varepsilon )=-\frac{i\, {\rm sign }\, \varepsilon }2
\left( \frac1{\tau }+\frac1{\tilde{\tau }}\; \sigma _z \right) ,
\end{equation}
where
\begin{equation}
\label{11}
\frac1{\tau }=\pi N_i\, V_0^2
\left( \nu _+ +\nu _- \right) ,
\end{equation}
\begin{equation}
\label{12}
\frac1{\tilde{\tau }}
=\pi N_i\, V_0^2M
\left( \frac{\nu _+}{\lambda _+}
-\frac{\nu _-}{\lambda _-}\right) ,
\end{equation}
$\lambda _\pm \equiv \lambda (k_{F,\pm })$, and
\begin{eqnarray}
\label{13}
\nu _+ =\frac{m}{2\pi }
\left| 1-\frac{m\alpha ^2}{\lambda _+}\right| ^{-1},\hskip1cm
\nonumber \\
\nu _- =\theta (\mu -M)\; \frac{m}{2\pi }
\left| 1-\frac{m\alpha ^2}{\lambda _-}\right| ^{-1}
\end{eqnarray}
are the densities of states at the Fermi surfaces of two different
subbands for $\mu >-M$. Here $\theta (x)$ is the Heaviside step
function, and $k_{F,\pm}$ are the Fermi momenta of the majority
and minority electrons, respectively.

Taking into account the self-energy correction (10), we find
the Green function of the electron system with impurities
\begin{eqnarray}
\label{14}
G_{\bf k}(\varepsilon )
=\frac{\varepsilon +i\, \Gamma -\varepsilon _{\bf k}
+\mu +\alpha (k_y\sigma _x
-k_x\sigma _y)-\sigma _z(M+i\, \tilde{\Gamma })}
{\varepsilon -E_{{\bf k},+ }+\mu
+i\,{\rm sign}\, \varepsilon /2\tau _+ }
\nonumber \\
\times \,
\frac1{\varepsilon -E_{{\bf k},- }+\mu
+i\, {\rm sign}\, \varepsilon /2\tau _- }
\; ,\hskip0.5cm
\end{eqnarray}
where
\begin{equation}
\label{15}
\Gamma =\frac{{\rm sign}\, \varepsilon }{2\tau }\; ,\hskip0.2cm
\tilde{\Gamma }=\frac{{\rm sign} \, \varepsilon }{2\tilde{\tau }}\; ,
\hskip0.2cm
\frac1{\tau _{\pm }}=\frac1{\tau }\,
\pm \, \frac1{\tilde{\tau }}\; \frac{M}{\lambda _\pm }.
\end{equation}
The values of $\tau _+ $ and $\tau _- $ determined by Eq.~(15) are the
relaxation times of electrons in different subbands.

\section{Vertex correction}

The equation for the renormalized vertex $\mathcal{T}(\varepsilon ,\omega )$
can be presented using the
diagrams with the impurity ladder included into the vertex part.\cite{agd}
For a short-range impurity potential, this
equation has the following form
\begin{equation}
\label{16}
\mathcal{T}_i(\varepsilon ,\omega )=
v_i+N_i\, V_0^2
\int \frac{d^2{\bf k}}{(2\pi )^2}\;
G_{\bf k}(\varepsilon )\,
\mathcal{T}_i(\varepsilon ,\omega )\, G_{\bf k}(\varepsilon +\omega ).
\end{equation}
In the limit of $\omega \to 0$, the integral is not zero only
at the Fermi surface, i.e., for energies
$\varepsilon \ll \tau _{\uparrow ,\downarrow }^{-1}$.
We assume that the density of impurities is low, which corresponds to
the large relaxation times $\tau _{\uparrow ,\downarrow }$.
Thus, we calculate the vertex renormalization for $\varepsilon =0$,
otherwise we take $\mathcal{T}_x=v_x$.

Denoting $\mathcal{T}_x=\mathcal{T}_x(\varepsilon =0,\, \omega \to 0)$, we
look for a solution of Eq.~(16) in the form
\begin{equation}
\label{17}
\mathcal{T}_x=ak_x+b\sigma _x+c\sigma _y
\end{equation}
with some coefficients $a$, $b$, and $c$.
After substituting this expression into (16) and using Eq.~(14), we find that in
the limit of low impurity density, $a=1/m$ and $b=0$, and the constant $c$
is defined by
\begin{eqnarray}
\label{18}
c\, \left[ 1-N_i\, V_0^2\int \frac{d^2{\bf k}}{(2\pi )^2}\;
\frac{(\mu -\varepsilon _{\bf k})^2-M^2}{D_+\, D_-}\right] \hskip0.5cm
\nonumber \\
=-\alpha \left[ 1+2N_i\, V_0^2
\int \frac{d^2{\bf k}}{(2\pi )^2}\;
\frac{\varepsilon _{\bf k}\, (\mu -\varepsilon _{\bf k})}{D_+\, D_-}
\right] \; ,
\end{eqnarray}
where $D_\pm=(\mu-E_{{\bf k},+}\pm i/2\tau _+)\, (\mu-E_{{\bf k},-}\pm i/2\tau _-)$.

Using Eqs.~(2), (6), (17) and (18), we conclude that the vertex renormalization
in Eq.~(6) for $\sigma _{xy}$
results in a substitution of the velocity
$v_x$ by $\mathcal{T}_x=k_x/m-\alpha ^*\sigma _y$ where
\begin{eqnarray}
\label{19}
\alpha ^*=\alpha \,
\left[ 1+2N_i\, V_0^2
\int \frac{d^2{\bf k}}{(2\pi )^2}\;
\frac{\varepsilon _{\bf k}\, (\mu -\varepsilon _{\bf k})}{D_+\, D_-}\right]
\hskip0.5cm \nonumber \\ \times
\left[ 1-N_i\, V_0^2\int \frac{d^2{\bf k}}{(2\pi )^2}\;
\frac{(\mu -\varepsilon _{\bf k})^2-M^2}{D_+\, D_-}\right] ^{-1}.
\end{eqnarray}
We should emphasize that the vertex renormalization refers only to the
states near the Fermi surface.

The integrals in Eq.~(19) can be calculated by transforming them to
integrals
over $\varepsilon _{\bf k}$, i.e., $\int d^2{\bf k}\, (2\pi )^{-2}...
=\int \nu _0(\varepsilon _{\bf k})\, d\varepsilon _{\bf k}...$,
where $\nu _0(\varepsilon )=(m/2\pi )\, \theta (\varepsilon )$.
Finally, we find
\begin{eqnarray}
\label{20}
\alpha ^*=\alpha \,
\left\{ 1+\frac{2\pi N_i\, V_0^2}{(\mathcal{E}_1-\mathcal{E}_2)^2}
\left[
\nu _0(\mathcal{E}_1)\;
\frac{\mathcal{E}_1\, (\mu -\mathcal{E}_1)}{\Gamma _1}
\right. \right. \nonumber \\ \left. \left.
+\, \nu _0(\mathcal{E}_2)\;
\frac{\mathcal{E}_2\, (\mu -\mathcal{E}_2)}{\Gamma _2}
\right] \right\}
\nonumber \\ \times
\left\{
1-\frac{\pi N_i\, V_0^2}{(\mathcal{E}_1-\mathcal{E}_2)^2}
\left[
\nu _0(\mathcal{E}_1)\;
\frac{\left[ (\mu -\mathcal{E}_1)^2-M^2\right] }{\Gamma _1}
\right. \right. \nonumber \\ \left. \left.
+\, \nu _0(\mathcal{E}_2)\;
\frac{\left[ (\mu -\mathcal{E}_2)^2-M^2\right] }{\Gamma _2}
\right] \right\}^{-1},
\end{eqnarray}
where
\begin{equation}
\label{21}
\mathcal{E}_{1,2}=\mu+m\alpha ^2
\pm \left( m^2\alpha ^4+2\mu m\alpha ^2+M^2\right) ^{1/2}
\end{equation}
and
\begin{eqnarray}
\label{22}
\Gamma _{1,2}
=\mp \,
\frac{\mu -\mathcal{E}_{1,2}
-\left( M^2+2m\alpha ^2\mathcal{E}_{1,2}\right) ^{1/2}}
{2\tau _+\, (\mathcal{E}_1-\mathcal{E}_2)}
\nonumber \\
\mp \,
\frac{\mu -\mathcal{E}_{1,2}
+\left( M^2+2m\alpha ^2\mathcal{E}_{1,2}\right) ^{1/2}}
{2\tau _-\, (\mathcal{E}_1-\mathcal{E}_2)}\; .
\end{eqnarray}
If $\mu >M$ (Fermi level is located within two subbands like shown in Fig.~1)
and in the limit of $\alpha \to 0$, we can find from (22) that
$\alpha ^*/\alpha \to 0$.

In the case of $-M<\mu <M$ (only the lowest energy subband is partly
filled with electrons) and $\alpha \to 0$, we obtain
\begin{equation}
\label{23}
\frac{\alpha ^*}{\alpha }=1-\frac{\mu +M}{2M}\; ,
\end{equation}
which gives us $\alpha ^*/\alpha \to 1$ for $\mu \to -M$, and
$\alpha ^*/\alpha \to 0$ for $\mu \to M$. The dependence of
$\alpha ^*/\alpha $ on $\alpha $ for different values of $\mu $
is shown in Fig.~2.

\begin{figure}
\vspace*{-1cm}
\hspace*{-0.5cm}
\includegraphics[scale=0.52]{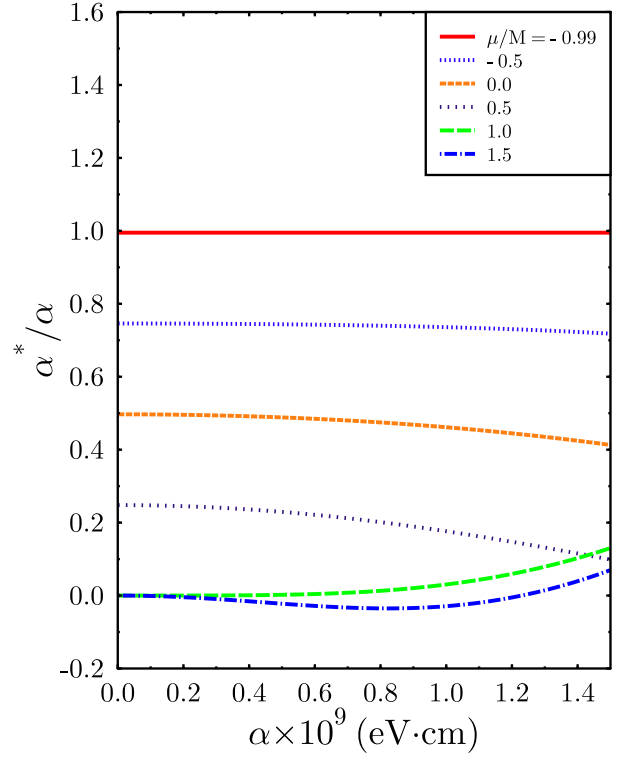}
\vspace*{-0.5cm}
\caption{Renormalized constant of SO
interaction $\alpha ^*$ in the vertex $\mathcal{T}_{x}$ at the
Fermi surface for different locations of the Fermi level.}
\end{figure}

\section{Anomalous Hall effect and topology of the energy bands}

The integration over energy $\varepsilon $ in Eq.~(6) with Green
function (14) leads to the separation of $\sigma _{xy}$ into two
parts, $\sigma _{xy}=\sigma _{xy}^{I}+\sigma ^{II}_{xy}$,
corresponding to the above-mentioned contribution of energy states
at the Fermi surface and the states below the Fermi energy,
respectively.\cite{streda82,crepieux01}

First we calculate the contribution from the states below the
Fermi energy, $\sigma ^{II}_{xy}$. Using (2), (6) and (14), we
find $\sigma ^{II}_{xy}$ as an integral over momentum with the
Fermi-Dirac functions $f(E_{{\bf k},\pm })$. Thus, it accounts for
the contribution of all states with $E_{{\bf k},+ }<\mu $ and
$E_{{\bf k},- }<\mu $:
\begin{eqnarray}
\label{24}
\sigma ^{II}_{xy}(\omega )
=-\frac{2ie^2}{\omega }\, {\rm Tr}
\int \frac{d^2{\bf k}}{(2\pi )^2}\,
\frac{f(E_{{\bf k},+ })-f(E_{{\bf k},- })}
{(E_{{\bf k},+ }-E_{{\bf k},- })^3}
\nonumber \\
\times \left( \frac{k_x}{m}-\alpha \, \sigma _y\right)
\left( E_{{\bf k},+ }-\varepsilon _{\bf k}
+\alpha k_y\sigma _x
-\alpha k_x\sigma _y
\right. \nonumber \\ \left.
-M\sigma _z \right)
\left( \frac{k_y}{m}+\alpha \, \sigma _x\right)
\left( -\omega +E_{{\bf k},+ }
-\varepsilon _{\bf k}
+\alpha k_y\sigma _x
\right. \nonumber \\ \left.
-\, \alpha k_x\sigma _y-M\sigma _z \right) .\hskip0.3cm
\end{eqnarray}
In the static limit of $\omega \rightarrow 0$, and after
calculating the trace, we obtain from (24)
\begin{equation}
\label{25}
\sigma ^{II}_{xy}
=-4e^2M\alpha ^2\,
\int \frac{d^2{\bf k}}{(2\pi )^2}\;
\frac{f(E_{{\bf k},+ })-f(E_{{\bf k},- })}
{(E_{{\bf k},+ }-E_{{\bf k},- })^3}\; .
\end{equation}
Using Eq.~(8) we can calculate the integral over momentum and obtain finally
\begin{equation}
\label{26}
\sigma ^{II}_{xy}
=\frac{e^2}{4\pi }\left[
1-\frac{M}{\lambda _+}
-\theta (\mu -M)\left( 1-\frac{M}{\lambda _-}\right)
\right] .
\end{equation}
In the limit of weak SO interaction, $\alpha k_{F,\pm }\ll M\, $, we
get from (26)
\begin{equation}
\label{27}
\sigma ^{II}_{xy}\simeq \frac{e^2}{2\pi }\;
\frac{m\alpha ^2}{M}
\left[ \theta (M-\mu )\; \frac{\mu +M}{2M}+\theta (\mu -M)\right] .
\end{equation}

The expression for $\sigma ^{II}_{xy}$ can be also presented in a
different form demonstrating the topological character of this
contribution.\cite{volovik88} Let us introduce a 3D unit vector
${\bf n}({\bf k})$ at each point of two-dimensional momentum plane
\begin{equation}
\label{28}
{\bf n}({\bf k})=\left(
\frac{\alpha k_y}{\lambda (k)}\; ,\;
-\, \frac{\alpha k_x}{\lambda (k) }\; ,\;
-\, \frac{M}{\lambda (k)}\right) .
\end{equation}
By using the ${\bf n}$-field (28), we parameterize the manifold of
$2\times 2$ Hermitian matrices corresponding to Hamiltonian (1),
since $H_0=\varepsilon _{\bf k}+\lambda (k)\, \bsig \cdot {\bf
n}({\bf k})$. At ${\bf k}=0$, the vector ${\bf n}$ is
perpendicular to the ${\bf k}$-plane, whereas for large $\left|
{\bf k}\right| \gg M/\alpha $, it lies in the ${\bf k}$-plane and
is oriented perpendicular to the momentum ${\bf k}$, like schematically
presented in Fig.~3. The dependence ${\bf n}({\bf k})$ is a
mapping of the ${\bf k}$-plane to the unit sphere $S_2$. As we can see
from Fig.~3, the total ${\bf k}$-plane maps onto the lower
hemisphere of $S_2$.

\begin{figure}
\hspace*{-0.7cm}
\includegraphics[scale=0.37]{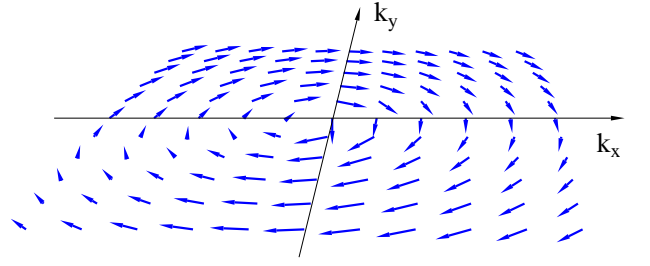}
\vspace*{-0.7cm}
\caption{The field ${\bf n}({\bf k})$ in the momentum
space.}
\end{figure}

Using (24) and (28) we find that in terms of the ${\bf n}$-field,
the contribution $\sigma ^{II}_{xy}$ can be written as
\begin{eqnarray}
\label{29}
\sigma ^{II}_{xy}
=-\frac{e^2}2
\int \frac{d^2{\bf k}}{(2\pi )^2}\;
\left[ f(E_{{\bf k},+ })-f(E_{{\bf k},-})\right]
\nonumber \\
\times \; \epsilon _{\alpha\beta\gamma}\, n_\alpha \;
\frac{\partial n_\beta }{\partial k_x}\;
\frac{\partial n_\gamma }{\partial k_y}\; ,
\end{eqnarray}
where $\epsilon _{\alpha\beta\gamma}$ is the unit antisymmetric tensor.
The integral
\begin{eqnarray}
\label{30}
\Omega =\frac12
\int \frac{d^2{\bf k}}{(2\pi )^2}\;
\left[ f(E_{{\bf k},+ })-f(E_{{\bf k},-})\right]
\nonumber \\ \times \;
\epsilon _{\alpha\beta\gamma}\, n_\alpha \;
\frac{\partial n_\beta }{\partial k_x}\;
\frac{\partial n_\gamma }{\partial k_y}
\end{eqnarray}
is the spherical angle on $S_2$ enclosed by two contours $L_+$
and $L_-$, where $L_\pm $ are the mappings of the Fermi surfaces
(circles) $E_{{\bf k}_F,+}=\mu $ and $E_{{\bf k}_F,-}=\mu $ to the
sphere $S_2$, respectively.

Notice that for $-M< \mu <M$, and in the case of $M/\alpha \ll
k_{F,+ }<2m\alpha \, (1+M/m\alpha ^2)^{1/2}$, we obtain $\Omega
\simeq 2\pi $, corresponding to the mapping of the ${\bf k}$-plane
to the whole unit hemisphere. This is possible only in the case of
large SO interaction or very small magnetization, when
$M\ll m\alpha ^2$.

There is also a contribution to the Hall conductivity from the
vicinity of Fermi surface. The calculation of corresponding term
using Eqs.~(6), (14) and (15) gives us
\begin{eqnarray}
\label{31}
\sigma ^{I}_{xy}
=-e^2M\alpha \alpha ^* \int \frac{d^2{\bf k}}{(2\pi )^2}\;
\frac1{(E_{{\bf k},+ }-E_{{\bf k},- })^2} \left[
\left( 1-\frac{\tau _+ }{\tau _- }\right) \right. \nonumber \\
\left. \times \left( -\frac{\partial f(E_{{\bf k},+ })}{\partial
\varepsilon }\right) +\left( 1-\frac{\tau _- }{\tau _+ }\right)
\left( -\frac{\partial f(E_{{\bf k},- })}{\partial \varepsilon
}\right) \right] .\hskip0.3cm
\end{eqnarray}
Here the presence of factor $(-\partial f/\partial \varepsilon )$
restricts (31) to the integral over Fermi surface (we consider
only the low-temperature limit of $T/(\mu +M)\ll 1$). Calculating
the integral over momentum (31), we find
\begin{eqnarray}
\label{32}
\sigma ^{I}_{xy}
=-\frac{e^2\alpha \alpha ^*M}4
\left[
\left( 1-\frac{\tau _+ }{\tau _- }\right)
\frac{\nu _+ }{\lambda ^2(k_{F,+ })}
\right. \nonumber \\ \left.
+\, \left( 1-\frac{\tau _- }{\tau _+ }\right)
\frac{\nu _- }{\lambda ^2(k_{F,- })}
\right] .
\end{eqnarray}
Using (11), (12) and (15), we can present $\sigma ^I_{xy}$ depending
only on the parameters of the energy spectrum and density of
states at the Fermi level
\begin{eqnarray}
\label{33}
\sigma ^{I}_{xy}
=-\frac{e^2\alpha \alpha ^*M^2\left( \xi _+-\xi_-\right) }2
\left\{
\frac{\xi _+ }{\lambda _+[\nu _++\nu_-+M(\xi _+-\xi _-)]}
\right. \nonumber \\ \left.
+\, \frac{\xi _- }{\lambda _-[\nu _++\nu_--M(\xi _+-\xi _-)]}
\right\} ,\hskip0.3cm
\end{eqnarray}
where $\xi _\pm=\nu _\pm/\lambda _\pm $. As we see from Eq.~(33),
the contribution of $\sigma ^I_{xy}$ does not depend on impurities
at all but is an intrinsic property of the crystal.

It follows from (33) that in the limit of large magnetization
(weak SO interaction), $M\gg \alpha k_{F,+ }$, the contribution of
this term is
\begin{equation}
\label{34}
\sigma ^I_{xy}
\simeq \frac{e^2}{4\pi }\; \frac{m\alpha \alpha ^*}{M}
\left[
\theta (M-\mu )\frac{m\alpha ^2}{M}
-4\, \theta (\mu -M)\left( \frac{m\alpha ^2}{M}\right) ^2
\right] .
\end{equation}

\begin{figure}
\vspace*{-1cm}
\hspace*{-0.5cm}
\includegraphics[scale=0.52]{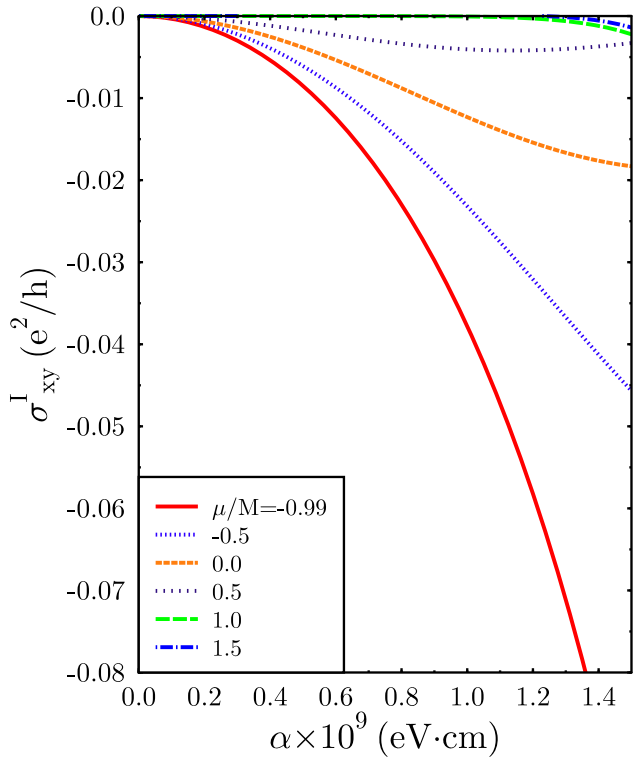}
\vspace*{-0.5cm}
\caption{$\sigma ^I_{xy}$ in units of $e^2/h$ as a function of SO coupling
constant $\alpha $ for different filling of the energy bands.}
\end{figure}

\begin{figure}
\vspace*{-1cm}
\hspace*{-0.5cm}
\includegraphics[scale=0.52]{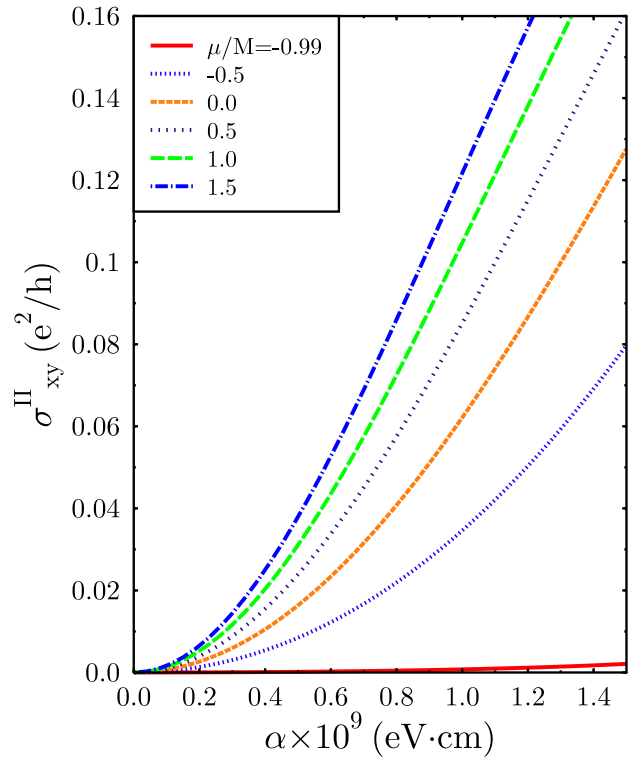}
\vspace*{-0.5cm}
\caption{$\sigma ^{II}_{xy}$ in units of $e^2/h$ as a function of $\alpha $
for different $\mu $.}
\end{figure}

\begin{figure}
\vspace*{-1cm}
\hspace*{-0.5cm}
\includegraphics[scale=0.52]{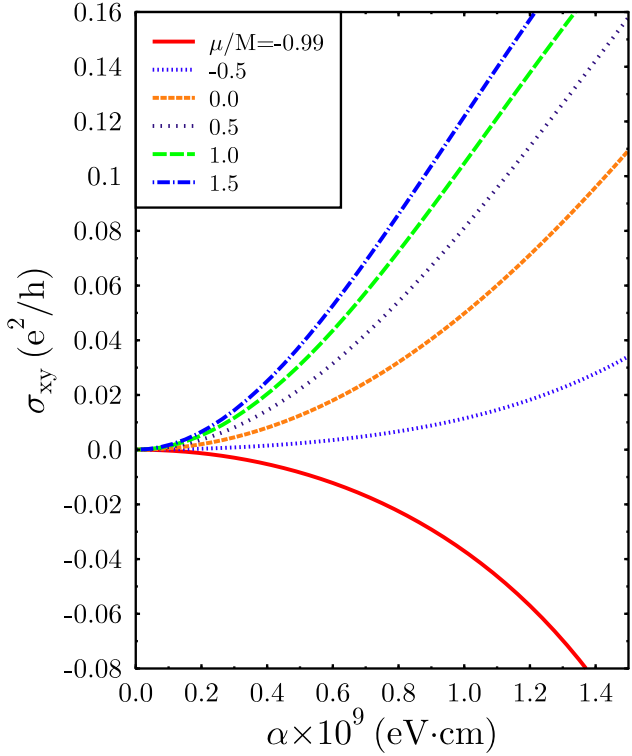}
\vspace*{-0.5cm}
\caption{Total Hall conductance $\sigma _{xy}=\sigma ^{I}_{xy}+\sigma ^{II}_{xy}$
in units of $e^2/h$ as a function of $\alpha $ for different $\mu $.}
\end{figure}

The dependence of two contributions to the Hall conductivity on
the magnitude of coupling constant $\alpha $ for different
positions of the Fermi level are presented in Figs.~4 and 5. We
calculated them using the parameters $M=0.01$~eV and $m=m_0$. Note
that the contribution of $\sigma ^{I}_{xy}$ does not depend on the
impurity density but only on the ratio between the relaxation
times for majority and minority carriers. The figures demonstrate
that $\sigma ^{I}_{xy}$ and $\sigma ^{II}_{xy}$ can be of the same
order of magnitude, and for $\mu <M$, the effect of $\sigma
^{I}_{xy}$ can dominate. The total Hall conductivity is presented
in Fig.~6. It shows that the AHE can change the sign with the
density of carriers.

The magnitude of SO coupling constant $\alpha $ can be roughly
estimated if we take the SO splitting in a semiconductor like
GaAs, $\Delta E_{SO}\simeq \alpha k=0.5$~meV corresponding to the
Fermi momentum $k=(2\pi N_s)^{1/2}$ for the carrier density
$N_s=10^{11}$~cm$^{-2}$ as a characteristic value for GaAs/GaAlAs
heterostructures.\cite{pfeffer95} It gives us $\alpha \simeq
6.3\times 10^{-10}$~eV$\cdot $cm. The experimentally obtained
magnitude of $\alpha $ in InGaAs/InAlAs heterostructures is
about $10^{-9}$~eV$\cdot $cm.\cite{nitta97}

\section{Separation of the contributions to Hall conductivity: general case}

Let us consider now a general case of the energy spectrum
described by a Hamiltonian $H_0$, which is a matrix in a certain
basis of wave functions. The off-diagonal conductivity can be
calculated using the Kubo formula
\begin{eqnarray}
\label{35}
\sigma _{xy}(\omega )=\frac{e^2}{\omega }
\int \frac{d\varepsilon }{2\pi}\,
\sum _{\bf k}\;
\sum _{nn^\prime mm^\prime }(v_x)_{nm}\hskip0.5cm
\nonumber \\
\times \,
G_{{\bf k}\, mm^\prime}(\varepsilon +\omega )\; (v_y)_{m^\prime n^\prime }\;
G_{{\bf k}\, n^\prime n}(\varepsilon )\, ,
\end{eqnarray}
where we can take the basis of eigenfunctions of the Hamiltonian
$H_0$ and assume a proper renormalization of velocities due to the
impurity
corrections. In this representation and in the absence of impurities, the
Green function $G^0_{\bf k}(\varepsilon )$ is diagonal. However, the
scattering from impurities induces nonzero off-diagonal elements,
as we can see
from the above-considered model (1) leading to the self energy with
nonvanishing off-diagonal matrix elements after diagonalization of
the Hamiltonian $H_0$.

The off-diagonal matrix elements of the Green function can be
formally presented in the following form resulting from an expansion
over off-diagonal self-energy matrix $\Sigma ^{(off)}$
\begin{eqnarray}
\label{36}
G_{{\bf k}\, nn^\prime }(\varepsilon )
\simeq G^{(0)}_{{\bf k}\, nn}(\varepsilon )\;
\Sigma ^{(off)}_{nn^\prime }(\varepsilon )\;
G^{(0)}_{{\bf k}\, n^\prime n^\prime }(\varepsilon )
\nonumber \\
=\frac{i\, {\rm sign}\, \varepsilon }{\tilde{\tau }_{nn^\prime }}\;
\frac{P_{nn^\prime }({\bf k})}{D_{\bf k}(\varepsilon )},\hskip0.5cm
n\ne n^\prime ,
\end{eqnarray}
where $\tilde{\tau }_{nn^\prime }$ is the relaxation time
associated with impurity-induced interband transitions,
$P_{nn^\prime }({\bf k})$ is a coefficient determined by the
specific form of Hamiltonian, $D_{\bf k}(\varepsilon ) =\prod
_n\left( \varepsilon -E_{{\bf k}n}+i\, {\rm sign}\, \varepsilon
/2\tau _n\right) $, $E_{{\bf k}n}$ is the dispersion of electrons
in the $n$-th subband, and $\tau _n$ is the corresponding
relaxation time, which includes all intra- and interband
transitions.

In the above-studied two-band model of Sec.~2, the
relaxation time $\tilde{\tau }_{12}$ equals to $\tilde{\tau }$ of
Eq.~(12), and $P_{12}({\bf k})=\alpha k/\lambda $. It should be
emphasized that both $\tilde{\tau }_{nn^\prime }$ and $\tau _n$
are inversely proportional to the density of impurities and to the matrix
elements of scattering from impurities.

We consider first the terms in Eq.~(35) with diagonal elements of
the Green functions (i.e., $m=m^\prime $ and $n=n^\prime $). In the static
limit of $\omega \to 0$, the nonvanishing contribution comes from
the states close to the Fermi surface, belonging to the same
energy branch $E_{{\bf k}n}$ (in other words, we should also take
$n=m$). Then we find
\begin{equation}
\label{37}
\sigma _{xy}^{IA}=e^2\sum _n\sum _{\bf k}
\left( -\frac{\partial f(E_{{\bf k}n})}{\partial \varepsilon }\right)
(v_x)_{nn}\, (v_y)_{nn}\, \tau _n\, .
\end{equation}
This contribution to the Hall conductivity is proportional to the
relaxation time of electrons. Correspondingly, it has the form of
the skew scattering mechanism.\cite{crepieux01,dugaev01}

Let us consider now the contribution from terms with $n=m$ and
$n^\prime =m^\prime $ in Eq.~(35). For $\omega \to 0$ and using
(36) we find
\begin{eqnarray}
\label{38}
\sigma _{xy}^{IB}=e^2\sum _{n\ne n^\prime }
\sum _{\bf k}
\left( -\frac{\partial f(E_{{\bf k}n})}{\partial \varepsilon }\right)
(v_x)_{nn}\, (v_y)_{n^\prime n^\prime }
\nonumber \\ \times
\frac{\tau _n}{\tilde{\tau }_{nn^\prime }}\;
\frac{P_{nn^\prime }({\bf k})}{\prod _{m\, (\ne n)}
\left( E_{{\bf k}n}-E_{{\bf k}m }\right) }
\end{eqnarray}
As follows from Eq.~(38), this contribution does not depend on the
impurity density and on the scattering amplitude because both
inverse relaxation
times $1/\tau _n$ and $1/\tilde{\tau }_{nn^\prime }$ are proportional to
the same impurity density and scattering amplitude.
Then the ratio $\tau _n/\tilde{\tau }_{nn^\prime }$ can be presented
using only the band parameters like
in the case of model in Sec.~2, see Eq.~(33). Hence, the contribution
(38) should be also identified as an intrinsic mechanism even though it
was calculated using the impurity
scattering. In this result we recognize the well-known property of
the side-jump mechanism.\cite{crepieux01} The only difference from the earlier
considered models is that the SO interaction is included here into
the Hamiltonian of free electrons instead of the impurity potential.

Coming back to Eq.~(35) with $n\ne m$ and $n^\prime \ne m^\prime$,
we find one more contribution to the AHE, which has been discovered
recently\cite{jungwirth02,yao04,m_onoda02} and also called {\it intrinsic}
mechanism of the AHE
\begin{eqnarray}
\label{39}
\sigma ^{II}_{xy}(\omega )=\frac{e^2}{\omega }
\int \frac{d\varepsilon }{2\pi}\, \sum _{\bf k}\;
\sum _{n\ne m} (v_x)_{nm}
\nonumber \\
\times \, G_{{\bf k}\, mm}(\varepsilon +\omega )\;
(v_y)_{mn}\; G_{{\bf k}\, nn}(\varepsilon )\, ,
\end{eqnarray}
where we omit some small corrections related to the off-diagonal
elements of the Green functions. After integrating over energy,
this result can be presented in the following
form\cite{jungwirth02,yao04,m_onoda02}
\begin{equation}
\label{40}
\sigma _{xy}^{II}=e^2\sum _n\sum _{\bf k}\; f(E_{{\bf
k}n})\, \left( \frac{\partial A_y({\bf k}n)}{\partial k_x}
-\frac{\partial A_x({\bf k}n)}{\partial k_y}\right) ,
\end{equation}
where
\begin{equation}
\label{41}
A_\alpha ({\bf k}n)=-i\left< {\bf k}n\left|
\frac{\partial }{\partial k_\alpha } \right|{\bf k}n\right>
\end{equation}
is the gauge potential in the momentum space related to the
transformation of the
Hamiltonian $H_0$ to the diagonal form. In a general case, this
transformation is local in the ${\bf k}$-space, leading to the
nonvanishing gauge potential $A_\alpha ({\bf k}n)$.

In the model of 2DEG with Rashba Hamiltonian we can calculate
explicitly the eigenfunctions
\begin{equation}
\label{42}
\left< {\bf k},\pm \right|
=\sqrt{\frac{\lambda (k)\pm M}{2\lambda (k)}}
\left(
1\, ,\; -\frac{i\alpha (k_x-ik_y)}{M\pm \lambda (k)}
\right) .
\end{equation}
Then using (41) we find the gauge potential
\begin{equation}
\label{43}
{\bf A}({\bf k},\pm )=\left(
-\frac{\alpha ^2k_y}{2\lambda (k)\, [M\pm \lambda (k)]}\; ,\;
\frac{\alpha ^2k_x}{2\lambda (k)\, [M\pm \lambda (k)]}\right) ,
\end{equation}
and from Eq.~(39) we come again to the same result of Eq.~(25).
Note that (43) can be also found as a gauge potential
corresponding to the local transformation of vector field
(28) to the homogenous field oriented along axis $z$ in the
momentum space (like in the case of local transformations in
the real space\cite{bruno04}).

In the 2D case, the flux of curl of the gauge potential
${\bf A}({\bf k},\pm )$ in Eq.~(40) through the surface $E_{{\bf
k},\pm }=\mu $ can be presented as the circulation of vector ${\bf
A}({\bf k},\pm )$ along the circle (Fermi surface in 2D). In other
words, the contribution of the filled states below the Fermi
surface can be also presented by the integral of the gauge field
over the Fermi surface. Recently, it was shown by Haldane
\cite{haldane04} that such a reduction of the integral in
momentum space takes place in any dimensionality, in accordance
with the Landau concept of the Fermi liquid stating that the
transport properties are fully determined by the properties of
electrons near the Fermi surface.

\section{Conclusions}

We calculated the off-diagonal conductivity using a model of 2DEG
with Rashba-type SO interaction. Starting from the Kubo formalism,
we found two contributions to the AHE, one of which, $\sigma
_{xy}^{II}$ presents the "intrinsic" mechanism and is related to
the topology of electron energy bands in the momentum space. The
other contribution, $\sigma _{xy}^I$, is related to impurities,
and it is not vanishing in the limit of small impurity density. We
demonstrate that both contributions can be of the same order of
magnitude.

As we see from Eqs.~(25) and (31), the Hall conductivity
$\sigma _{xy}$ is proportional
to the second order of SO coupling $\alpha $. The even order in
$\alpha $ is related to the form of SO interaction in Eq.~(1), for
which the space inversion, ${\bf k}\to -{\bf k}$ is equivalent to
the change of sign $\alpha \to -\alpha $. Indeed, the nonvanishing
integral over momentum in Eq.~(6) requires the integrand to be
invariant with respect to ${\bf k}\to -{\bf k}$, which excludes
any odd in $\alpha $ factors in $\sigma _{xy}$.

It is worth to note that the impurity-related mechanisms of AHE
(both side-jump and skew scattering) lead to the non-vanishing off-diagonal
conductivity in the first order of $\alpha $.\cite{crepieux01}
The SO interaction included in the impurity scattering potential has the
form of $igV_{{\bf kk}^\prime }\bsig \left( {\bf k}\times {\bf k}^\prime \right) $,
where $g$ is a constant and $V_{{\bf kk}^\prime }$ is the matrix
element of impurity potential. This type of SO interaction is
invariant with respect to spatial inversion.
Thus, in the case of small SO interaction, the side-jump or skew scattering
can be dominating as they both are of the first order of SO interaction.

The essential point of our consideration is that we do not neglect the impurity
scattering leading to a finite relaxation time of electrons at the Fermi
surface. Hence, the transition to the "clean limit" in our approach means
that we should take the limit of small density of impurities only
in the final formula for conductivity after the limit of $\omega \to 0$.
The opposite way of calculation would lead us to break of continuous
transition between the "clean limit" and any small impurity density.
This point was completely neglected in the previous consideration of
the intrinsic mechanism of AHE. Fortunately, it does not affect the
"geometric ingredient" of AHE, $\sigma ^{II}_{xy}$, but, as we
demonstrated in this paper, it is crucial for the competing contribution
$\sigma ^I_{xy}$.

{\it Note added:} After completing the manuscript we were informed
that similar results have been also obtained using a different method by
Sinitsyn, Niu, Sinova and Nomura.\cite{sinitsyn05} We thank Jairo Sinova for
this information.  

\begin{acknowledgments}
V.D. thanks the University Joseph Fourier and Laboratory Louis
N\'eel (CNRS) in Grenoble for kind hospitality. This work was
supported by FCT Grant No.~POCTI/FIS/58746/2004 in Portugal,
Polish State Committee for Scientific Research under Grants
PBZ/KBN/044/P03/2001 and 2~P03B~053~25, and also by Calouste
Gulbenkian Foundation.
\end{acknowledgments}

\end{document}